\documentclass{article}
\usepackage{amsmath}
\usepackage{amssymb} 
\usepackage{graphicx}
\usepackage{authblk}
\numberwithin{equation}{section}
\newcommand{\keywords}[1]{\textbf{\textit{Index terms---}} #1}

\begin{document}

\title{A Study on the time resolution of Glass RPC}
\author[1,2]{ N. Dash}
\author[1,2]{V. M. Datar}
\author[3]{ G. Majumder}

\affil[1]{Nuclear Physics Division, Bhabha Atomic Research Centre, Mumbai - 400085, INDIA}
\affil[2]{Homi Bhabha National Institute, Anushaktinagar, Mumbai - 400094, INDIA}
\affil[3]{Tata Institute of Fundamental Research, Mumbai - 400005, INDIA}

\maketitle

\begin{abstract}
The 50~kton Iron Calorimeter (ICAL) detector at the underground India based Neutrino Observatory (INO) will make measurements on atmospheric neutrinos. Muons produced in charged current (CC) interactions of muon neutrinos with the iron are tracked spatially and temporally through the signals that they produce in the Resistive Plate Chambers~(RPCs) that are interleaved with iron layers. Since the RPCs will be operated in the avalanche mode the signal rise-time is $\sim~1~\rm{nsec}$ resulting in  a fast time response. While the muon track is derived from the X and Y hit information of the RPCs and the layer number (Z), the upward or downward direction is obtained by using the time information from the detector. Such a capability can be examined by analysing the timing information from $1~\rm{m}~\times~1~\rm{m}$ glass RPCs, with $3~\rm{cm}$ wide X- and Y- pick-up strips, in a $12$ layer RPC stack that measures cosmic muon events. The present study looks at the pixel-wise time response of these RPCs in order to improve the relative time distribution and hence the up-down discrimination capability. After including the effect of propagation delay in the cable and pick-up panel the time resolution improves, in some cases, to $\leq~1~\rm{nsec}$ whereas in some cases there is no significant change. These results will help in significantly improving on the extraction of the directionality of muons produced in CC interactions of $\nu_{\mu}$ and $\bar{\nu}_{\mu}$.
\end{abstract}


\section{Introduction}
Exploration of unresolved problems in the atmospheric neutrino sector is the principal goal of the proposed Iron Calorimeter (ICAL) \cite{ical} detector at the India-based Neutrino Observatory (INO) \cite{ino}. Apart from making accurate measurements of neutrino mixing parameters using muon neutrinos and anti-neutrinos, separately, it will address the mass hierarchy problem. The atmospheric muon neutrinos ($\nu_{\mu}$) interact with nucleons through charged current (CC) and neutral current (NC) interactions. In the CC interaction of a neutrino with a target nucleon the final state has the corresponding lepton and hadrons but in a NC interaction the in-elastically scattered neutrino is accompanied by hadrons. Due to the presence of a neutrino in the final state of NC interaction this process is difficult to use for getting information about the incoming neutrino. On the other hand in the CC interaction with a muon in the final state the energy and direction of the incoming neutrino can be extracted by combining the energy and momenta of the muon and the hadrons. As presented in Fig.1 \cite{nuflux1} the flux of the atmospheric neutrinos is 10 times higher up to 10~GeV than below it at the INO site. So ICAL is mainly looking for neutrino with energy up to 10~GeV for physics studies. By considering the energy loss of muon in 56~mm thick iron plate, typically a muon of energy from $0.8~\rm{GeV}$ to few TeV will be able to measured by the ICAL detector.

The ICAL detector will consist of 3 modules, each weighing about 17~ktons and having dimensions of $\sim$ $16~\rm{m}~\times~16~\rm{m}~\times~15~\rm{m}$. The detector will consist of alternate layers of soft iron of thickness 56~mm and glass Resistive Plate Chambers (RPCs) about $2~\rm{m}~\times~2~\rm{m}$ in dimensions placed in inter-layer gaps of 40~mm. The RPC gives X- and Y- information corresponding to the ionization produced by any charged particle, such as a muon, passing through the detector. This is derived from the electrical signal induced by the passage of the charged particle, due to the multiplication of charge on account of the high electric field applied between the anode and cathode, on the pick-up strips of $\sim$3~cm width placed on either side of RPC and orthogonal to each other. Since the RPC will be operated in the avalanche mode the signal rise-time is $\sim$1~nsec. In a CC interaction of the muon neutrino the outgoing muon track is derived from the X and Y hit information of the RPCs and the upward or downward direction is obtained by using the time information from the detector. Typically the time difference $\Delta \rm{t}$ between consecutive RPC layers of $\sim$9.6~cm and by considering the zenith angle from $0^0~-~75^0$ is between $0.32$ to $1.24~\rm{nsec}$. In general 5-3 minimum number of RPC layers and with strip multiplicities 1-3 for consecutive strips are used to generate a trigger. In this paper we report the results of a study of the pixel-wise time response of the 1~m$^2$ glass RPCs to seek a possible improvement in time resolution. By correcting for the effect of propagation delay in the cable and in the pick-up panel the time resolution is improved to $\leq$ 1~nsec. However in a few cases there is no significant change. These results will help in significantly improving on the extraction of the directionality of muons produced in CC interactions of $\nu_{\mu}$ and $\bar{\nu}_{\mu}$. 

\begin{figure}
\centering
\includegraphics[width=0.5\textwidth]{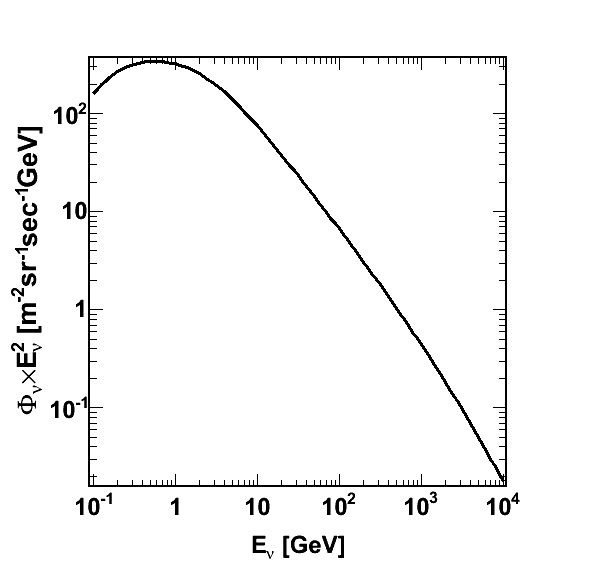}
\caption{The atmospheric neutrino flux averaged over all directions and summed over $\nu_e$ + $\bar{\nu}_e$ + $\nu_{\mu}$ + $\bar{\nu}_{\mu}$ as a function of neutrino energy at the INO site.}
\end{figure}

\section{Description of Prototype Detector}
The RPC is a gaseous detector composed of two parallel electrodes made of float glass with a volume resistivity of about $10^{12}$ $\Omega.\rm{cm}$. The two  electrodes of 3~mm thickness, are held 2~mm apart by means of suitable spacers. High voltage is applied to the outer surfaces of the glass plates which are coated with graphite having an area resistivity of $\sim$ 1 $\frac{M\Omega}{\square}$. The gas mixture of Freon (95.15$\%$), Iso-Butane (4.51$\%$) and $SF_6$ (0.34$\%$), and the suitable applied voltage enables the detector to operate in the avalanche mode. As the electrode is of high resistance glass the spark discharges within a limited area. A stack consisting of 12 RPC detectors of dimensions ($1~\rm{m}~\times~1~\rm{m}$) installed at TIFR has been working for more than 10 years. Each 1~m~$\times$~1~m RPC has 32 parallel pick-up strips on each side laid orthogonally. These X, Y signal strips map the zone of interaction in a RPC detector. The detector efficiencies are above $95\%$ \cite{eff} as measured using cosmic-ray muons. Apart from assessing the long term performance of RPCs, it has also been used to make measurements on the cosmic muon flux \cite{eff}. Such detectors of $2~\rm{m}~\times~2~\rm{m}$ size will be used in large numbers in the flagship experiment to make measurements on atmospheric muon neutrinos using a magnetised Iron Calorimeter at the India-based Neutrino Observatory.

The information one needs from the stack are the event time, co-ordinates of the interaction zone, its relative time of interaction in the detector with respect to event trigger and the status of the detector. Due to the avalanche mode of operation of the detector the signal strength is of the order of few mV. The necessary amplification of the signal, from a pick-up strip has been done using a two stage cascading amplifier with a gain of $\sim$80. The amplified output is fed to an analog discriminator cum analog front end (AFE) with a voltage threshold at -20~mV. Then the AFE output is fed to digital front end board (DFE). Each individual strip signal is stretched to 720~nsec as the trigger signal arriving at DFE is delayed due to the longer signal path. The pre-trigger signals from AFE are used to form level~1 signals which are further used for generation of an event trigger in a CAMAC based Data Acquisition (DAQ) system. To generate a triggered event, these level 1 ECL signals from different layers are fed to a coincidence module via an ECL to NIM converter module. Data in the DFE board is latched and the TDC data is also stored for every trigger. The scalar is used to monitor the strip count rate irrespective of the trigger signal. All required control signals for event and monitoring  processes routed to DFEs via the Control and Data router~(CDR) board are initiated by the Control Module.  Event data is transferred serially to the Readout board. The DAQ system has been optimised for the VME format and the details are presented in Ref.\cite{daq}.


\section{Detector Time Resolution}
The detector time resolution is one of the criteria to judge the quality of a detector with respect to its response to minimum ionizing charged particles. So the time resolution describes quantitatively how precisely the time at which a particle crossed the detector can be determined. In this paper the time distribution is characterised by a Gaussian with the parameters sigma($\sigma$) centered around a time $t_{0}$, where the time distribution is parametrized as $\propto$ $e^{\frac{-(t-t_0)^2}{2\sigma^2}}$.

The overall time resolution of a detector arises from,\\
1. the fluctuation or spread in the time response of the detector,\\
2. contribution from the electronics (pre-amplifier and readout system which include time-walk and time-jitter), and\\
3. the time measurement device such as time to digital converter(TDC).

For a detector with high signal-to-noise ratio the time jitter is much smaller than the rise time, where as due to the different amplitudes there is a shift in the timing signal due to the use of leading edge discriminator termed as time-walk. This difference in amplitude arises from the avalanche process as well as the variation in attenuation of the signal due to the position dependent signal propagation path length through the pick-up panel depending on the point of induction in the detector. Figure 2 shows a schematic diagram of the detector. Figure 2 (left) shows possible hits of muons triggering a single X-strip at 3 locations in Y giving rise to 3 signal times $t_1$, $t_2$ and $t_3$. They have different propagation delays corresponding to their propagation distances (see Sec. 4).
\begin{figure}[here]
\centering
\begin{minipage}{.5\textwidth}
  \centering
  \includegraphics[width=40mm]{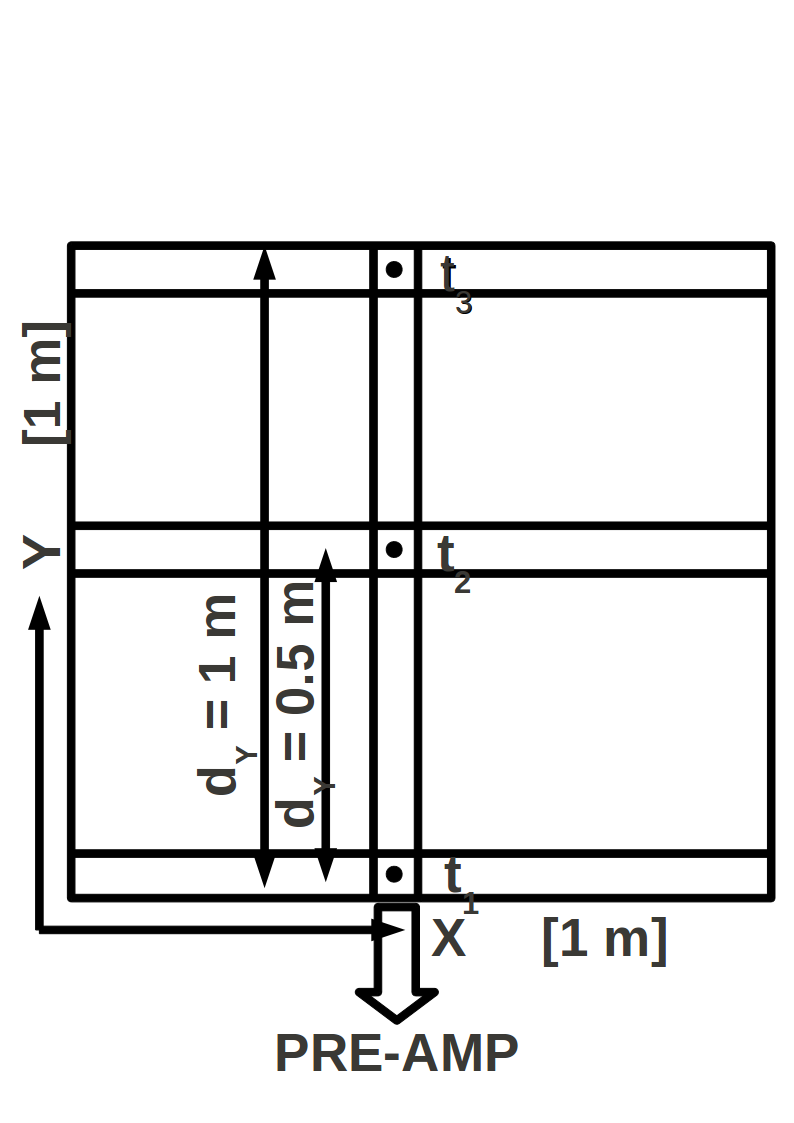}
  \label{fig:neff-sim}
\end{minipage}%
\begin{minipage}{.5\textwidth}
  \centering
  \includegraphics[width=55mm]{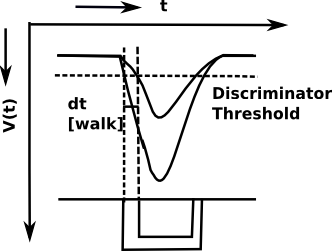}
  \label{fig:test2}
\end{minipage}
\caption{Position of hit point in a RPC (left), schematic diagram for voltage pulses showing time-walk of the discriminator output (right).}
\end{figure}

The overall relative time difference ($\Delta \rm{t}$) between two consecutive detectors (layer 6 and 7) is represented in Fig. 3. It is obtained with irrespective of the position of the hit point in the detector which results in a long tail \cite{total}. The value of $\sigma$/$\sqrt(2)$ of the distribution of the time difference will give the time resolution of a detector.

\begin{figure}[here]
\begin{center}
\includegraphics[width=0.5\textwidth]{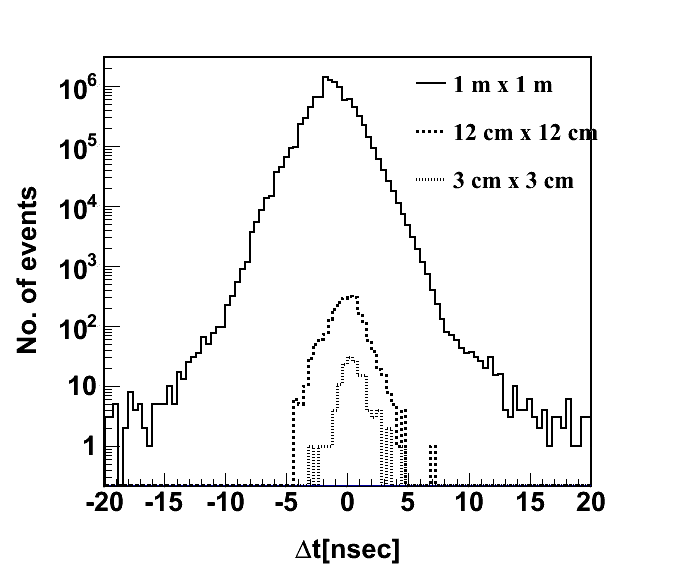}
\caption{Overall relative time difference of full area, $12~\times~12~\rm{cm}^2$ pixel and that for $3~\times~3~\rm{cm}^2$ pixel. The FWHM improves from 3.53 to 1.66 nsec.} \label{fig:eresp-n}
\end{center}
\end{figure}

In order to see the uniformity of this distribution of the time difference between two detectors, we have introduced ``Pixel Method''. Each 1$\times$1~$\rm{m}^2$ RPC has 32 strips per plane. In ``Pixel Method'' 32 strips are divided in to 8 groups. So each group has 4 strips. A similar scheme has been employed for the Y-plane. So there are a total of 64 groups with each of dimensions 12~cm~$\times$~12~cm or 4~$\times$~4~pixels. The divided area including both X and Y planes is shown in Fig. 4 (left) and the zoomed part of a single block is shown in Fig. 4 (right).
\begin{figure}[here]
\centering
\begin{minipage}{.5\textwidth}
  \centering
  \includegraphics[width=40mm]{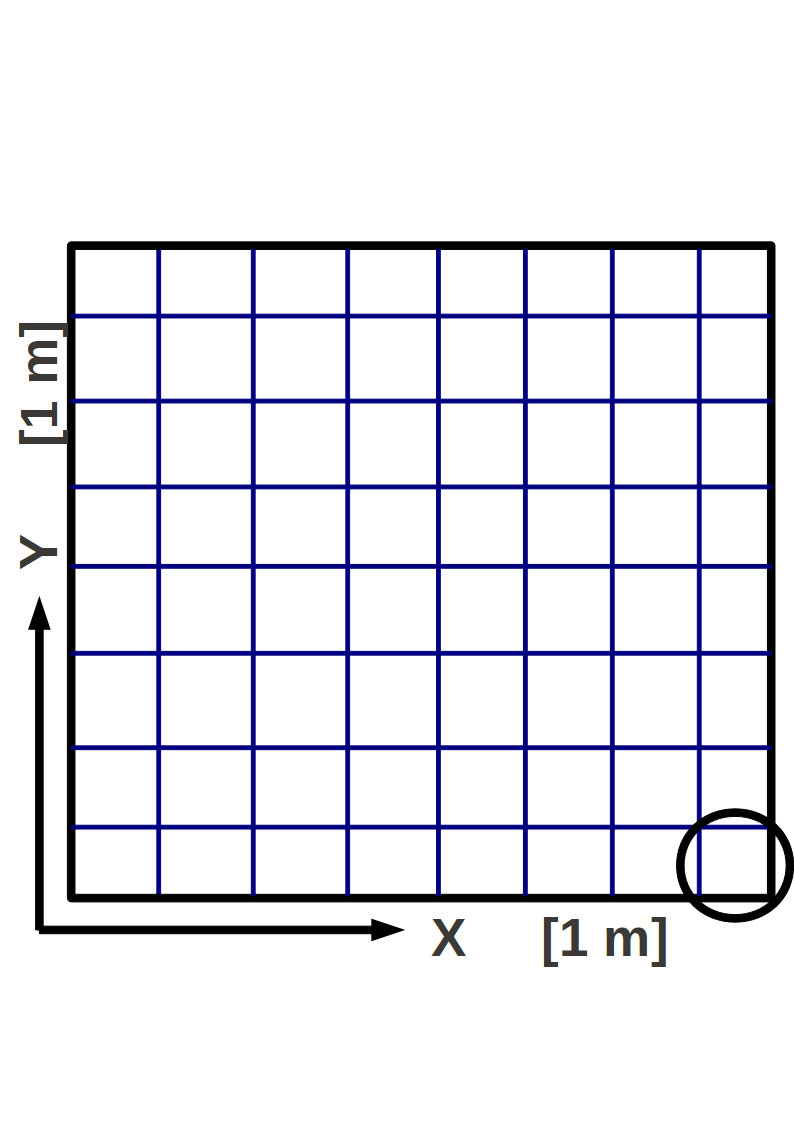}
  \label{fig:neff-sim}
\end{minipage}%
\begin{minipage}{.5\textwidth}
  \centering
  \includegraphics[width=40mm]{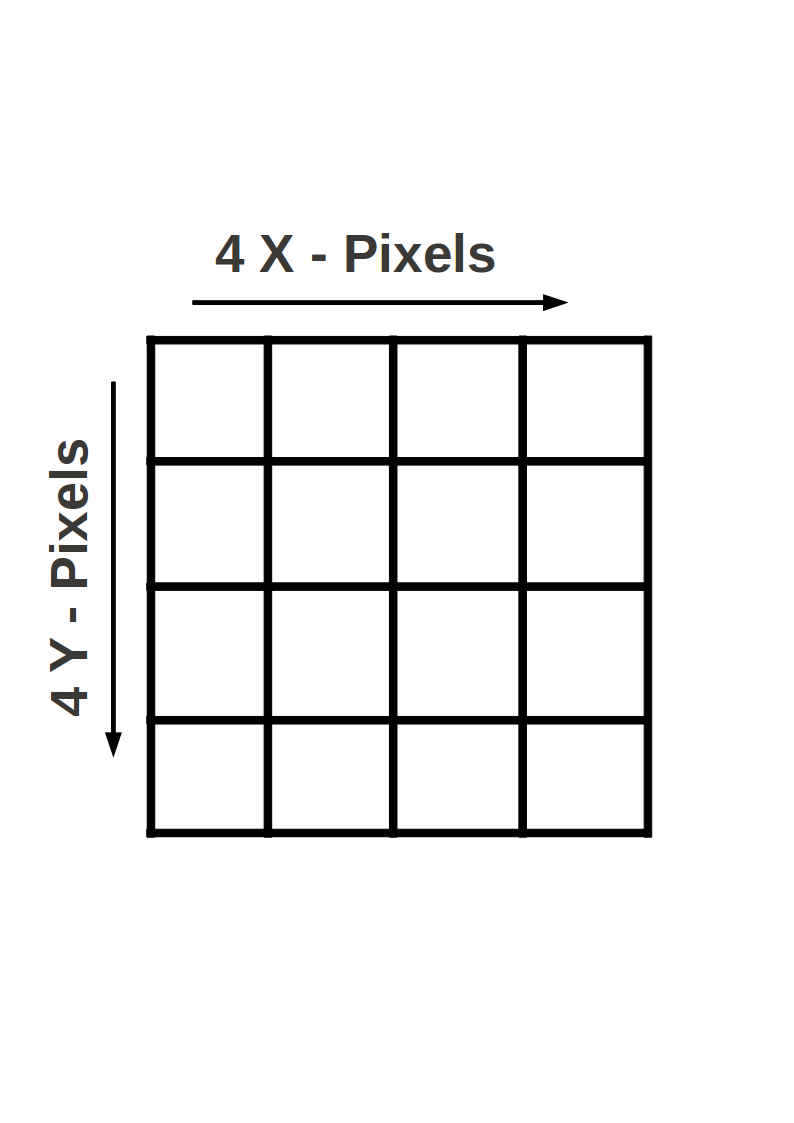}
  \label{fig:test2}
\end{minipage}
\caption{Grouping of RPC by Pixel Method with 64 pixels of size $12~\rm{cm}~\times~12~\rm{cm}$ (left), single pixel (right) of size $12~\rm{cm}~\times~12~\rm{cm}$.}
\end{figure}

A calculation of the relative time difference for each group has been done by considering the X and  Y co-ordinates of the muon hit position in that pixel. The difference in time of same detector layers for a pixel at the central region of area $12~\times~12~\rm{cm}^2$ and the corresponding single pixel with area $3~\times~3~\rm{cm}^2$ are shown in Fig. 3. 


\section{Prototype Data Analysis by Pixel Method using Cosmic-Ray Muons}
The analysis involves the processing of raw detector data. The raw data consist of the digitized output of detector electronic signals. Signals are induced in the detector electronics by the passage of cosmic-ray muons, which leave ``hits'' in active detector elements. So the raw data contains the X,Y co-ordinates of hits in terms of strip number and their corresponding time information. In the analysis as mentioned in Fig. 4, each detector is divided into 64 groups considering X and Y planes. Events having a hit in the triggered layer are considered for the analysis. The relative time difference of a pixel is obtained by considering the events with both X and Y position in that pixel. For this calculation electronic offset and the propagation delay due to pick-up panel are taken into account.
\begin{equation}
\rm{T_{hit} = T_{tdc} - T_{offset}}.
\end{equation}
and
\begin{equation}
\rm{T_{offset} = T_{eloffset} + T_{stripoffset}}.
\end{equation}
Where $\rm{T_{hit}}$ and $\rm{T_{tdc}}$ are respectively the time information of the hit position and the TDC. $\rm{T_{eloffset}}$ is the delay due to electronics for each strip on the pick-up panel. It is obtained by using cosmic ray muon data. $\rm{T_{stripoffset}}$ is obtained by taking the ratio of distance traveled by the signal through the pick-up panel and its propagation velocity, assumed to be equal to 2/3 of the velocity of light.
\begin{equation*}
\rm{T_{stripoffset} = \frac{\mathrm{d}[cm]}{\mathrm{v}[cm/sec]}},
\end{equation*}
where $\mathrm{v}$ is the velocity of propagation, and $\mathrm{d}$ is the distance travel.

After correcting for all these offsets the time difference between two layers is plotted by considering the time information of X-plane i.e. X tdc time information. Figure 5 and Fig. 6 shows the standard deviation of the distribution of relative time differences in $\it{nsec}$ per pixel between two consecutive layers from layer 0 to layer 11. The values of colour coding are represented by side bar near to each plot. These plots shows the homogeneity of the distribution over 1~m~$\times$~1~m RPC detector area. Some of the edge pixels have no events due to low detector acceptance, an artifact of trigger logic. The sum of the distribution of standard deviation for all these pixels (i.e. for 11$\times$64 pixels) is shown in Fig. 6 (last plot on the right side). For a pixel of 12~cm~$\times$~12~cm dimensions the minimum value of the time resolution is 0.636~nsec, where as the time resolution is peaking around 1.27~nsec. The tail part corresponds to the statistical uncertainties of the distribution in pixel wise which results in broad sigma. 


\begin{figure}[h]
\centering
\begin{minipage}{.5\textwidth}
  \centering
  \includegraphics[width=70mm,height=6.5cm]{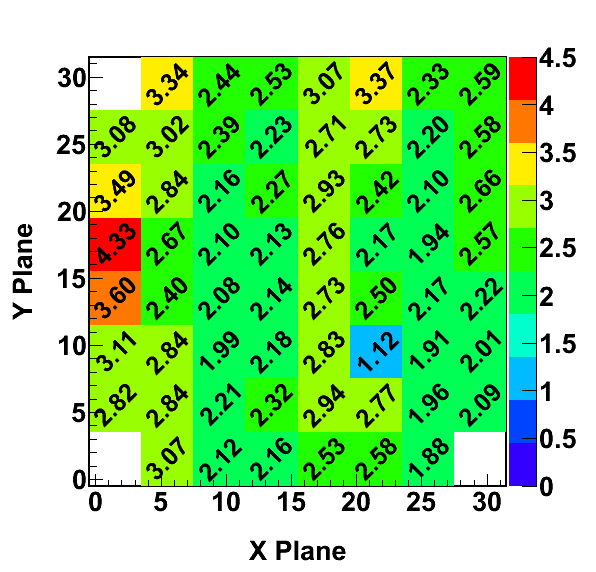}
  \label{fig1}
\end{minipage}%
\begin{minipage}{.5\textwidth}
  \centering
  \includegraphics[width=70mm,height=6.5cm]{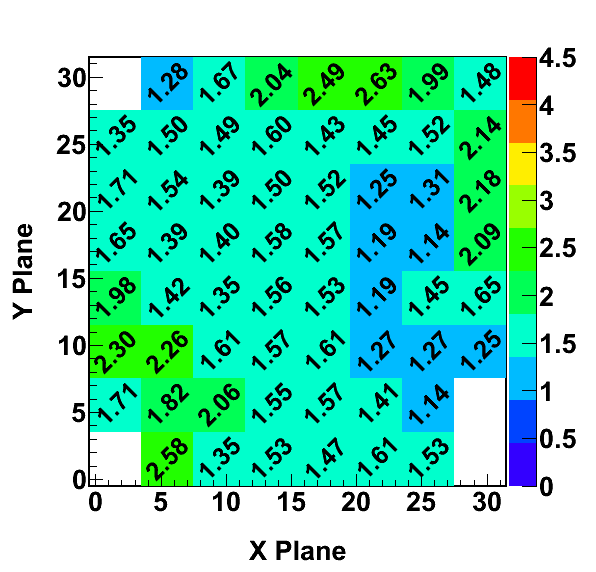}
  \label{fig2}
\end{minipage}\\
\begin{minipage}{.5\textwidth}
  \centering
  \includegraphics[width=70mm,height=6.5cm]{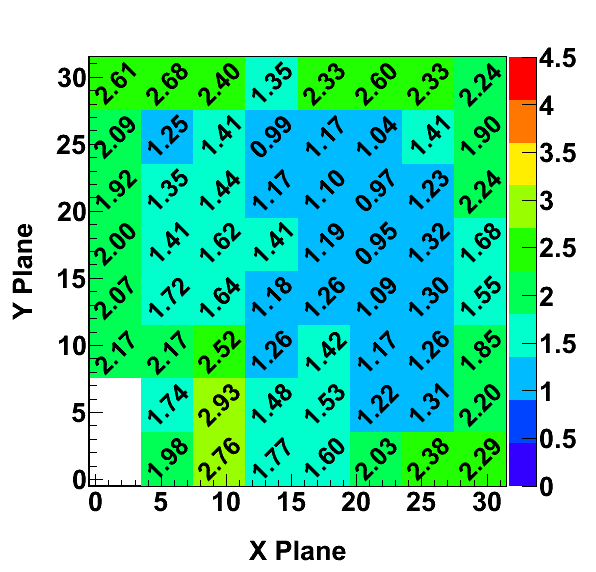}
  \label{fig3}
\end{minipage}%
\begin{minipage}{.5\textwidth}
  \centering
  \includegraphics[width=70mm,height=6.5cm]{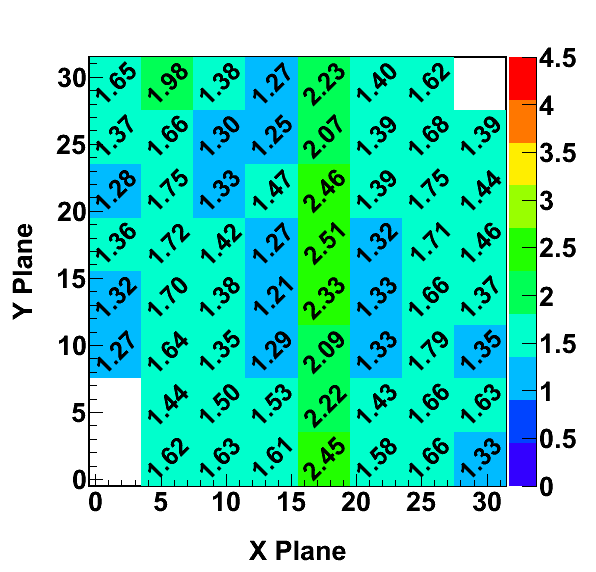}
  \label{fig4}
\end{minipage}\\
\begin{minipage}{.5\textwidth}
  \centering
  \includegraphics[width=70mm,height=6.5cm]{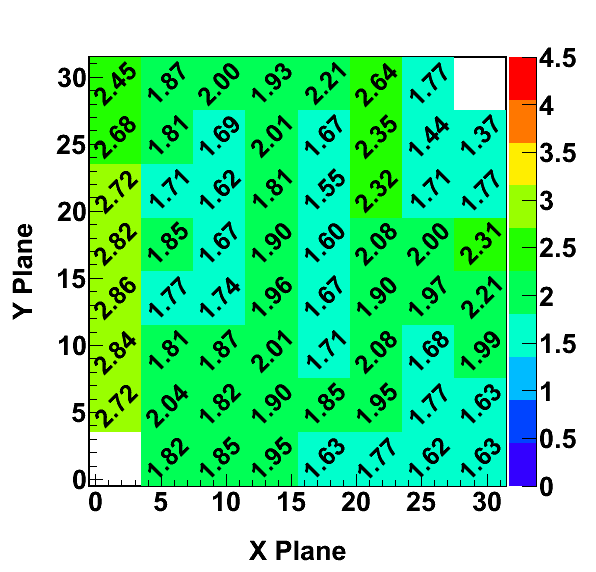}
  \label{fig5}
\end{minipage}%
\begin{minipage}{.5\textwidth}
  \centering
  \includegraphics[width=70mm,height=6.5cm]{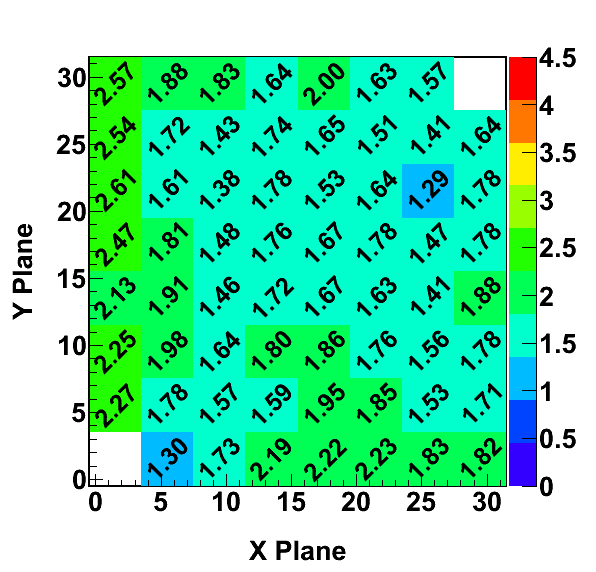}
\end{minipage}
\caption{The sigma of the relative time difference distribution obtained between two consecutive layers in row wise from left to right (Layer 0-1, 1-2, 2-3, 3-4, 4-5, 5-6)}
\end{figure}

\begin{figure}[h]
\centering
\begin{minipage}{.5\textwidth}
  \centering
  \includegraphics[width=70mm,height=6.5cm]{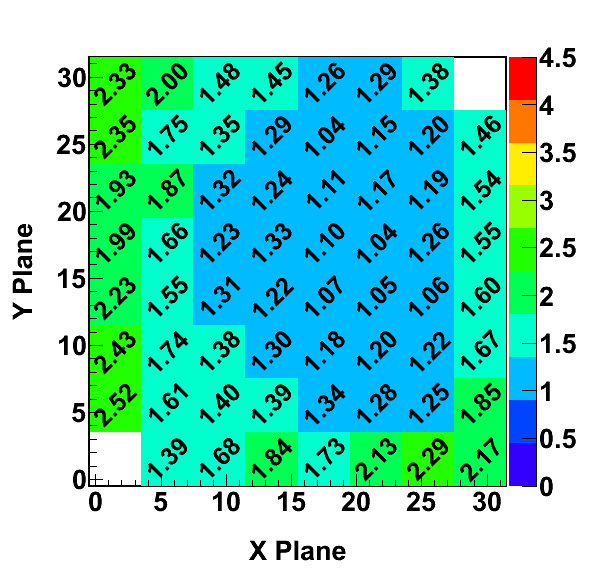}
  \label{fig:neff-sim}
\end{minipage}%
\begin{minipage}{.5\textwidth}
  \centering
  \includegraphics[width=70mm,height=6.5cm]{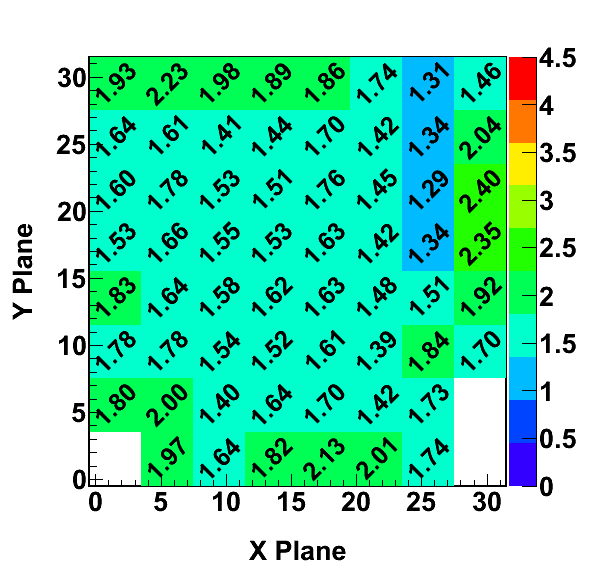}
  \label{fig:test2}
\end{minipage}\\
\begin{minipage}{.5\textwidth}
  \centering
  \includegraphics[width=70mm,height=6.5cm]{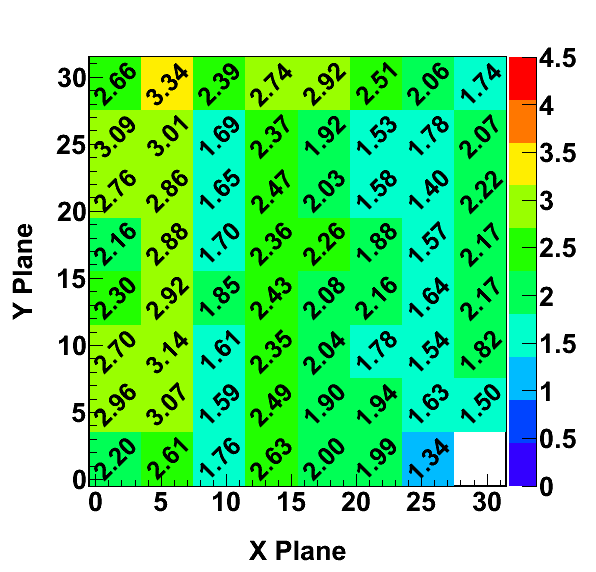}
  \label{fig:neff-sim}
\end{minipage}%
\begin{minipage}{.5\textwidth}
  \centering
  \includegraphics[width=70mm,height=6.5cm]{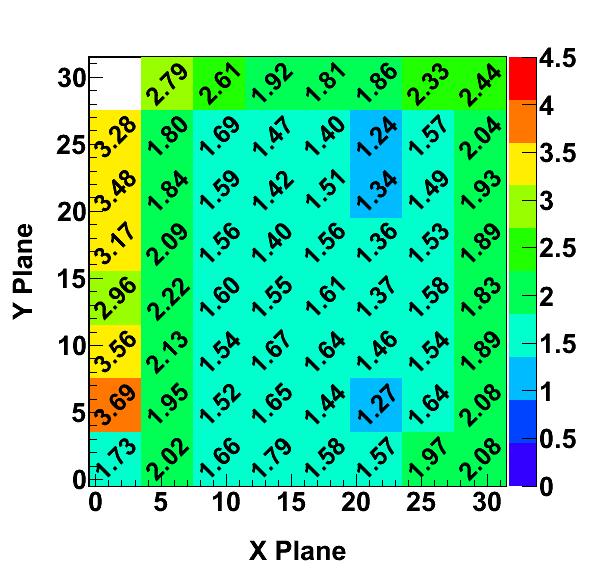}
  \label{fig:test2}
\end{minipage}\\
\begin{minipage}{.5\textwidth}
\centering
  \includegraphics[width=70mm,height=6.5cm]{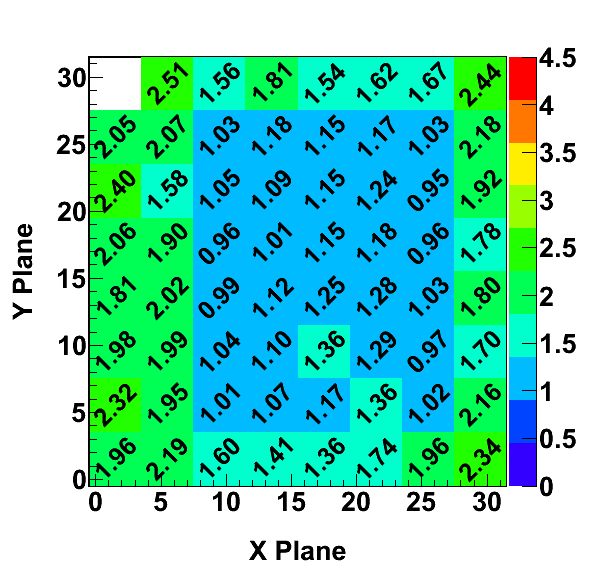}
  \label{fig:neff-sim}
  \end{minipage}%
  \begin{minipage}{.5\textwidth}
  \centering
  \includegraphics[width=80mm,height=6.5cm]{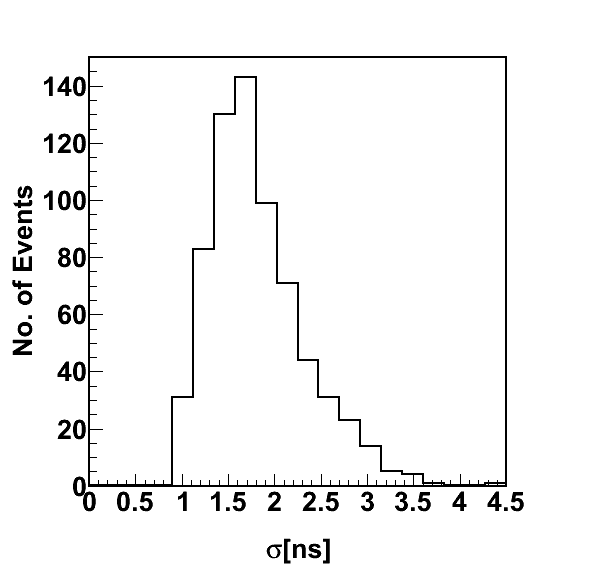}
  \label{fig:neff-sim}
  \end{minipage}
\caption{The sigma of the relative time difference distribution obtained between two consecutive layers in row wise from left to right (Layer 6-7, 7-8, 8-9, 9-10, 10-11). The last plot on the right side corresponds to the distribution of standard deviations for 11$\times$64 pixels.}
\end{figure}



\section{Summary}
In summary we have measured the time resolution of the glass RPC obtained by analysing prompt time spectra using cosmic ray muons incident on the 12 layers $1~\rm{m}^2$ RPC stack at TIFR, Mumbai. The overall time resolution improves when one considers $4~\times~4$ pixels with each pixel of size $\sim~3~\rm{cm}~\times~3~\rm{cm}$. This improves even further if one considers $1~\times~1$ pixels. In principle this will help in identifying the up-down directionality of the muons produced in neutrino interactions in the 50~kton ICAL-INO detector.


\noindent {\bf Acknowledgements}\\
The authors would like to thank Prof. Naba K. Mondal for his interest in, and support for, the work reported in this paper. Our thanks are also to Prof. B. S. Acharya and Prof. S. Saha for their critical comments, suggestions during the preparation of the manuscript. One of us (ND) would like to thank the India-based Neutrino Observatory, India for financial support.


\end{document}